\documentclass[prb,aps,epsf,twocolumn,showpacs,pstricks]{revtex4-1}
\usepackage{graphicx}
\usepackage{epsfig}
\usepackage{latexsym}
\usepackage{amsmath,amsfonts,amssymb}
\usepackage{color}
\usepackage{array}

\newcommand{\e}{\varepsilon}
\newcommand{\up}{\uparrow}
\newcommand{\down}{\downarrow}
\renewcommand{\>}{\rangle}
\renewcommand{\(}{\left(}
\renewcommand{\)}{\right)}
\renewcommand{\[}{\left[}
\renewcommand{\]}{\right]}
\renewcommand{\v}[1]{\mathbf{#1}} 

\begin{document}
\title{Engineering a p+ip Superconductor: Comparison of Topological Insulator and Rashba Spin-Orbit Coupled Materials}
\author{Andrew C. Potter and Patrick A. Lee}
\affiliation{ Department of Physics, Massachusetts Institute of
Technology, Cambridge, Massachusetts 02139}

\begin{abstract} We compare topological insulator materials and Rashba coupled surfaces as candidates for engineering p+ip superconductivity.  Specifically, in each type of material we examine 1) the limitations to inducing superconductivity by proximity to an ordinary s-wave superconductor, and 2) the robustness of the resulting superconductivity against disorder.  We find that topological insulators have strong advantages in both regards: there are no fundamental barriers to inducing superconductivity, and the induced superconductivity is immune to disorder.  In contrast, for Rashba coupled quantum wires or surface states, the the achievable gap from induced superconductivity is limited unless the Rashba coupling is large.  Furthermore, for small Rashba coupling the induced superconductivity is strongly susceptible to disorder.  These features pose serious difficulties for realizing p+ip superconductors in semiconductor materials due to their weak spin-orbit coupling, and suggest the need to seek alternatives.  Some candidate materials are discussed.
\end{abstract}
\maketitle

\section{Introduction}
Superconductors with $p+ip$ pairing symmetry have long been expected to posses zero-energy Majorana bound states in vortex cores\cite{Ivanov} or at the ends of one-dimensional structures \cite{Kitaev}.  These Majorana fermion bound states are expected to exhibit non-Abelian exchange statistics\cite{Ivanov,Read/Green}, and have been proposed as a basis for topological quantum computers which would be protected from decoherence\cite{Kitaev,Nayak,Alicea1DWires}.  Consequently, there is a growing interest in realizing a robust $p+ip$ superconductors in the laboratory. Such $p+ip$ superconductors are thought to naturally occur in triplet paired fermionic superfluids (such as $^3\text{He}$ A  or $\text{Sr}_2\text{RuO}_4$)\cite{He3Ref,Sr2RuO4Ref}, and in the Pfaffian quantum Hall state\cite{Moore/Read} at $\nu=5/2$.  However, these systems are all experimentally delicate, and despite extensive experimental work, direct evidence of Majorana fermions remains elusive. 

Recently, the possibility of engineering effective $p+ip$ superconductors in more conventional materials has arisen\cite{Fujimoto,Sau,AliceaSingle,Potter1,Potter2,PALee,Fu/Kane,QuantumWires,HalfMetals}.   
A common thread in these proposals is the use of spin-orbit coupling to convert conventional superconductivity into $p+ip$ superconductivity, typically by inducing s-wave superconductivity in a 2D material with spin-orbit coupling by proximity to an ordinary bulk  superconductor.  Among these proposals, two dominant classes of candidate materials have emerged: 1) surface states of topological insulator (TI) materials\cite{Fu/Kane} and 2) semiconducting quantum-wires or two-dimensional electron gases (2DEGs) with Rashba spin-orbit coupling $U_R$ and induced magnetization $V_z$\cite{Fujimoto,Sau,AliceaSingle}. In this work, we provide a detailed comparison of induced superconductivity in these two classes of materials and discuss the comparative advantages and disadvantages of using each of these materials to construct a $p+ip$ superconductor.  

We first examine the prospects for inducing superconductivity in TI surface states and Rashba materials by the proximity effect.  For TI surface states, the induced s-wave pairing is always  converted into $p+ip$ pairing due to the topologically protected winding of the TI surface Bloch wave-functions.  Consequently, for a sufficiently good interface between the TI surface and a bulk superconductor, it is possible to induce the full bulk pairing gap $\Delta_0$ on the TI surface\cite{TIInducedSC}.  The situation is more complicated for Rashba materials, where a more delicate balance of spin-orbit and magnetization is required to achieve p+ip superconductivity.  For these materials, the size of the induced superconducting gap is limited not only by the transparency of the interface to the bulk superconductor, but also by the magnetization and Rashba energy scales which we denote by $V_z$ and $U_R$ respectively (see Section \ref{sec:Rashba2DEG} for detailed definitions).  In particular for small Rashba coupling ($U_R\ll V_z,\Delta_0$), we find that the induced superconducting gap is limited to $\frac{1}{2}\sqrt{U_R\Delta_0}\ll\Delta_0$.

We then analyze the effects of disorder on the induced superconductivity. Since superconductivity in the 2D surface layer is induced by proximity rather than spontaneously developed by phonon interaction, and since the induced superconductivity has s-wave symmetry one might expect that the disorder cannot reduce the induced pairing gap.  On the other hand, disorder is pair-breaking for p-wave superconductors, and the induced superconductivity is effectively converted into p+ip superconductivity.  Therefore it is not a priori clear what the effect of disorder will be.  By computing the disorder averaged density of states, we find that, for TI materials, the induced superconductivity is immune to disorder, and argue on general grounds that this immunity is a direct consequence of time-reversal invariance.  In the Rashba 2DEG's however, the induced magnetization required to realize a single helicity $p+ip$ superconductor inherently breaks time-reversal symmetry leaving the induced superconductivity vulnerable to disorder.  We find that, while the induced superconducting gap $\Delta$ never fully closes from disorder, it can be sharply reduced from its clean value.  The degree of vulnerability to disorder depends again on the size of the Rashba coupling $U_R$.  For small $U_R$, $\Delta$ is strongly suppressed even for very weak disorder for which the superconducting coherence length $\xi_0$ is only a few percent of the mean-free path $\ell$.  This sharp decrease is more drastic than the case of magnetic impurities in a conventional superconductor, for which superconductivity is destroyed only when $\xi_0/\ell \sim 1$.

In both regards, the TI materials offer advantages, allowing robust induced superconductivity that is immune to disorder.  This suggests that TI materials may therefore be the most promising route to realizing topological superconductivity and Majorana fermions.  However, implementing a $p+ip$ superconductor using a TI surface state requires many further developements in material growth and interface engineering, and therefore it may still be desirable to work with more conventional materials with strong Rashba splitting.  

So far, the theoretical and preliminary experimental work on building a $p+ip$ superconductor from Rashba 2DEGs has largely focused on semiconductor materials and in particular on semiconductor nanowires\cite{QuantumWires}.  However, in light of our analysis, the very low Rashba energy scales in semiconductors raise serious challenges for inducing superconductivity.  Namely, small $U_R$ greatly limits the size of the induced superconducting gap and furthermore, renders the resulting superconductor extremely sensitive to disorder. These drawbacks suggest that an alternative class of materials with stronger spin-orbit coupling should be sought.  

Extremely large Rashba splittings on the order of $1$eV have been observed in surface alloys of metals and heavy elements such as Bi on Ag(111)\cite{SurfaceAlloy}.  In an earlier paper\cite{Potter1}, we had proposed this surface alloy as a promising candidate material. However, this extreme Rashba strength can also create new problems.  Namely, large Rashba coupling leads to large carrier density, making it difficult to adjust the chemical potential by  gating.  This is problematic because one must be able to fine-tune the chemical potential to achieve topological superconductivity and to manipulate Majorana end-states.  It is therefore desireable to find materials with strong enough Rashba couplings to avoid problems with induced superconductivity and disorder, but not so strong that gating becomes impossible.  

One particularly promising candidate is the (110) surface of Au, which first-principles calculations predict will exhibit surface bands with sizeable Rashba splitting\cite{AuSurface}.  These surface bands naturally lie within $\lesssim 50$meV of the bulk Fermi level, indicating that it should be possible to move the chemical potential into the topological regime using gating.  Furthermore, the crystal symmetry of the (110) surface allows for a combination of Rashba and Dresselhaus type spin-orbit couplings.  If both types of spin orbit coupling are present, magnetization could be induced by applying an in-plane field rather than by proximity to a ferromagnetic insulator\cite{AliceaSingle}.  This would greatly simplify the proposed setup for realizing Majoranas. 

This paper is organized as follows: we begin with a review of the proposed route to engineering a $p+ip$ superconductor from TI and Rashba 2DEG materials.  We then introduce a simple model of the proximity effect in these materials, and show how to choose system parameters in order optimize the induced superconducting gap.  Subsequently, we turn to the issue of disorder, and derive the disorder averaged Green's functions and density of states for weak to moderate disorder ($k_F\ell\gg 1$).  Finally, we close with a discussion of the relative strengths and weaknesses of each class of materials for realizing a $p+ip$ superconductor.

\section{Overview of Proposed Routes to Topological Superconductivity}
In this section we briefly review the proposed routes to creating a topological superconductor from the surface state of a bulk topological insulator (TI) or from a 2DEG with strong Rashba spin-orbit coupling.  In both classes of materials, spin-orbit coupling creates surface bands in which electron spin is locked with respect to the direction of propagation.  This helical locking of electron spin direction to propagation directions causes the electron spin to wind as one traverses a loop around the Brillouin zone.  Consequently, if s-wave superconductivity is induced by proximity to a bulk superconductor, the helical winding of the surface Bloch wave-functions effectively converts the induced superonductivity into $p+ip$ superconductivity.  Specifically, when re-expressed in the basis of the surface bands, the induced s-wave pairing term takes the form of a p-wave pairing term\cite{Fu/Kane,AliceaSingle}.

Throughout the paper, we work in the basis of time-reversed pairs: 
\begin{eqnarray} \Psi_k &=& \begin{pmatrix}\psi_k\\ \mathcal{T}\psi_k \end{pmatrix} = \begin{pmatrix}\psi_k\\ -i\sigma_y\mathcal{K}\psi_k \end{pmatrix}
= \begin{pmatrix}\begin{pmatrix} c_{k,\up} \\ c_{k,\down} \end{pmatrix} \\ \begin{pmatrix}c^\dagger_{-k,\down} \\ -c^\dagger_{-k,\up}\end{pmatrix} \end{pmatrix} \label{eq:TRBasis}\end{eqnarray}
which is convenient for discussing superconductivity.  Here we take the usual representation $\mathcal{T} = -i\sigma_y\mathcal{K}$ of the time-reversal operator, where  $\{\sigma_{x,y,z}\}$ are Pauli matrices in the spin basis, and $\mathcal{K}$ denotes complex conjugation.  

We consider inducing the pairing term:
\begin{eqnarray} H_\Delta &=& \Delta\sum_k c_{k,\up}^\dagger c_{-k,\down}^\dagger +h.c. \nonumber \\ 
&=&\sum_k \Delta\psi_k^\dagger \(\mathcal{T}\psi_k\)^\dagger +h.c. = \sum_k \Psi_k^\dagger \Delta\tau_1 \Psi_k \label{eq:PairingTerm}\end{eqnarray} 
by proximity to an ordinary superconductor, where $\{\tau_{1,2,3}\}$ are Pauli matrices in the particle-hole basis.  Here we have chosen a gauge in which the pairing order parameter $\Delta$ is purely real (this is justified, as we are not presently concerned with situations where the superconducting phase is inhomogenous or fluctuating).

In proposals to realize Majorana fermions, there are two relevant energy scales protecting the coherence of information stored among the Majorana fermions.  The first is the (p-wave component of the) induced superconducting pairing gap, $\Delta$, which sets the energy scale for single particle excitations that can change the fermionic parity of a pair of Majoranas.  The second is the so-called mini-gap to localized excitations near each Majorana.  Since localized excitations cannot change non-local fermion parity, the mini-gap is important only when two Majoranas are brought close to each other for measurement purposes\cite{Akhmerov}.  For Majoranas realized as end-states, the mini-gap also scales linearly with the induced p-wave pairing gap\cite{Potter2}.  Therefore, in order to perform quantum coherent manipulations of Majorana fermions, it is important to achieve a robust pairing gap, and to work at temperatures much lower than this gap.

\subsection{Topological Insulators}
The low-energy continuum Hamiltonian, $H_{\text{TI}} = \sum_k\Psi_k^\dagger \mathcal{H}_{\text{TI}}(k)\Psi_k$ of a TI surface is\cite{Fu/Kane}: 
\begin{equation} \mathcal{H}_{\text{TI}} =  \(v\hat{z}\cdot\(\sigma\times\mathbf{k}\)-\mu\)\tau_3 \label{eq:HTI}\end{equation} 
where $v$ is the Dirac cone velocity. The surface eigenstates with energies $\pm v|k|$ form the upper and lower branches of a single Dirac cone with definite spin-helicity:
\begin{equation} c_{\pm} = \(\cfrac{c_{k,\up}\pm e^{-i\phi_k}c_{k,\down}}{\sqrt{2}}\) \label{eq:HelicalBasis} \end{equation}
where $\phi_k = \tan^{-1}\(k_x/k_y\)$.

When expressed in the $c_{\pm}$ basis, the pairing term (Eq. \ref{eq:PairingTerm}) takes the form of an ideal spinless $p+ip$ superconductor\cite{Fu/Kane}
\begin{equation} H_\Delta = \sum_k \Delta\( e^{\displaystyle i\phi_k}c^\dagger_{k,+}c^\dagger_{-k,+}+e^{\displaystyle-i\phi_k}c^\dagger_{k,-}c^\dagger_{-k,-}+h.c.\) \end{equation}

\subsection{Rashba 2DEG \label{sec:Rashba2DEG}}
The low-energy continuum Hamiltonian, $H_{\text{R}} = \sum_k\Psi_k^\dagger \mathcal{H}_{\text{R}}(k)\Psi_k$ of a Rashba coupled 2DEG with induced magnetization is:
\begin{equation} \displaystyle \mathcal{H}_{\text{R}}(\mathbf{k}) = \[\xi_k+\alpha_R \hat{z}\cdot\(\sigma\times\mathbf{k}\)\]\tau_3 + V_z\sigma_z  \label{eq:HR} \end{equation}
where $\xi_k = \frac{k^2}{2m}-\mu$ is the spin-independent dispersion, $\mu$ is the chemical potential, $U_R = 2m\alpha_R^2$ is the Rashba coupling strength, and $V_z$ is the induced Zeeman splitting responsible for the surface magnetization. 

\begin{figure}[ttt]
\begin{center}
\includegraphics[width=3in]{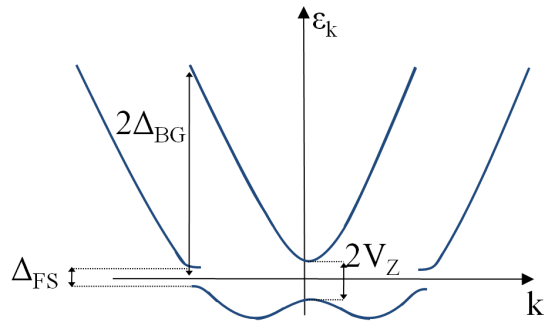}
\end{center}
\vspace{-0.2in}
\caption{(Color online) Band-structure with Rashba coupling, magnetization, and induced superconductivity.  Three relevant energy scales are labelled: $2\Delta_{\text{BG}}$ is the energy gap between the $\e_+$ and $\e_-$ bands at the Fermi surface, $\Delta_{\text{FS}}$ is the induced p-wave pairing gap at the Fermi surface, and $2V_Z$ is the induced magnetization gap.}
\label{fig:RashbaBS}
\end{figure}

In contrast to the TI case, without breaking time-reversal symmetry, there are two helicities present at each energy.  Consequently, in order to construct a single species $p+ip$ superconductor, it is necessary to explicitly break time-reversal symmetry, in this case by introducing magnetization, $V_z$, to remove one of these helicities. 

The Rashba coupling $\alpha_R$ creates two helical bands with energies $\e_{\pm}^{(R)} = \xi_k\pm \alpha_R|k|$ and spin-wavefunctions $c_{\pm}$ (as for the TI case).
$V_z$ cants the helical bands by angle $\theta_M(k)$ out of the xy-plane modifying the surface eigenstates and corresponding dispersions: 
\begin{eqnarray} &\e_{\pm}^{(\text{R/FM})} = \xi_k\pm\sqrt{V_z^2+U_R^2\frac{k^2}{2m}} \nonumber  \\&c_{\pm}^{(\text{R/FM})}= e^{\displaystyle{-i\phi_k\sigma_z/2}}e^{\displaystyle -i\theta_{M}\sigma_x/2}\(\cfrac{c_{k,\up}\pm c_{k,\down}}{\sqrt{2}}\)
\nonumber \\
&\theta_M(k)=\tan^{-1}\(V_z/\sqrt{V_z^2+U_R^2\frac{k^2}{2m}}\)\end{eqnarray}

Re-expressing $H_\Delta$ in the eigenbasis of both Rashba and Zeeman couplings, one finds that, in addition to $p\pm ip$ pairing $\Delta_{\text{p}}(k) \hat{\v{k}}^\pm\sim  \<c_{k,\pm}c_{-k,\pm}\>$ between fermions both in band $\e_{\pm}$, the canting $\theta_M$ introduces an s-wave pairing component $\Delta_{\text{s}}(k) \sim \<c_{k,+}c_{-k,-}\>$ between fermions $c_+$ and $c_-$ in bands $\e_{+}$ and $\e_{-}$ respectively where:
\begin{equation} \begin{pmatrix} \Delta_{\text{s}}(\v{k}) \\ \Delta_{\text{p}}(k) \end{pmatrix} = \frac{1}{2\sqrt{V_z^2+U_R^2\frac{k^2}{2m}}}\begin{pmatrix} V_z \\ -\sqrt{U_R^2\frac{k^2}{2m}} \end{pmatrix} \Delta \label{eq:DeltaSP}\end{equation}
and $\hat{\v{k}}^\pm = \(k_y\pm ik_x\)/k$.  
As discussed in \cite{AliceaSingle}, one has a topological superconductor with potential Majorana bound states so long as $V_z>\Delta$, and so long as $\mu$ lies within the Zeeman gap ($|\mu|<V_z$).  It is most advantageous to set $\mu=0$, placing the chemical potential in the middle of the Zeeman gap (which can be done either by electrostatic gating or chemical doping), and we will take $\mu=0$ throughout the remainder of this paper.

In this system, there are two excitation gap energy-scales: the first is the pairing gap at the Fermi surface ($k=k_F$) given by: 
\begin{eqnarray} &\displaystyle \Delta_{FS} = 2\Delta_p=\sqrt{\cfrac{U_R}{\Delta_{BG}}}\Delta \label{eq:DeltaFS}  \\
& \Delta_{BG} = \cfrac{\e_+^{(\text{R/FM})}-\e_-^{(\text{R/FM})}}{2} = \sqrt{V_z^2+U_R\cfrac{k_F^2}{2m}}  \end{eqnarray}
The second is the Zeeman gap at $k=0$, given (for $\mu=0$) by $|V_z-\Delta|$.  The smaller of these two energy scales sets the bulk gap to single-particle excitations which would destroy the non-local information stored among Majorana bound-states.  We note that the relative strength of the Rashba spin-orbit coupling $U_R$ and the Zeeman splitting $V_z$ determines the size of the pairing gap at $k_F$. For $V_z\gg U_R$, the pairing gap is only a small fraction of originally induced $\Delta$.   If the Zeeman gap closes (i.e. if $V_z\leq \Delta$), then both helicities are present and the system is topologically trivial.

\section{Proximity Induced Superconductivity \label{sec:InducedSC}}
\subsection{Simple Model of Proximity Effect}
In this section, we consider the interface between a bulk s-wave superconductor and either a topological insulator surface or a Rashba coupled surface state with induced magnetization $V_z$.  As a simple model of this interface, we consider a bulk superconductor described by the BCS Hamiltonian:
\begin{equation} H_{B} = \sum_{k,\sigma}\[ \e_{B,k}b_{k,\sigma}^\dagger b_{k,\sigma} + \(\Delta_0 b_{k,\up}^\dagger b_{-k,\down}^\dagger+h.c.\)\] \end{equation}
coupled to the surface through a clean planar interface described by the bulk--surface tunneling term:
\begin{equation} H_{\text{B--S}} = \displaystyle \sum_{\mathbf{k}_\parallel,k_\perp,\sigma}\Gamma b_{(\mathbf{k}_\parallel,k_\perp),\sigma}^\dagger c_{\mathbf{k}_\parallel,\sigma} +h.c.  \end{equation} 
which conserves momentum $\mathbf{k}_\parallel$ parallel to the interface, and is independent of the transverse momentum $k_\perp$ perpendicular to the interface.  Here $b_{k,\sigma}^\dagger$ and $c_{k,\sigma}^\dagger$ are the electron creation operators (with momentum $k$ and spin $\sigma$) for the bulk superconductor and surface respectively, $\e_B$ is the non-superconducting bulk dispersion which we will linearize about the chemical potential $\mu$, and $\Delta_0$ is the bulk s-wave pairing amplitude. 

Since surface--bulk tunneling conserves in-plane momentum, the bulk tunneling density of states (in the absence of superconductivity) is given by the one-dimensional expression $N_B(\e_B(k))= \(\partial \e_B(k)/\partial k_z\)^{-1}$.  Assuming that $N_B$ varies slowly with energy, $H_{\text{B--S}}$ induces the following self-energy correction to the surface Green's function:
\begin{equation} \Sigma_\Gamma(i\omega) = \frac{\pi\gamma}{\sqrt{\Delta_0^2+\omega^2}}\(-i\omega+\Delta_0\tau_1\)  \end{equation}
where $\gamma = N_B(0)|\Gamma|^2$ is convenient measure of the strength of surface-bulk coupling corresponding to the width of the surface resonance that would result from $H_{\text{B--S}}$ without bulk-superconductivity ($\Delta_0 = 0$).

Incorporating $\Sigma_\Gamma$ into the surface Green's function gives:
\begin{equation}\mathcal{G_S}(i\omega) = \frac{Z_\Gamma}{i\omega-Z_\Gamma\mathcal{H}_{\text{TI/R}}-(1-Z_\Gamma)\Delta_0\tau_1} \label{eq:SurfaceGrnsFn}\end{equation}
where $Z_\Gamma$ is the reduced in quasi-particle weight due to the bulk--surface hybridization:
\begin{equation} Z_\Gamma(i\omega) = \(1+\frac{\pi\gamma}{\sqrt{\Delta_0^2+\omega^2}}\)^{-1} \label{eq:SurfaceZ} \end{equation}
The quasi-particle weight can be interpreted as the fraction of time that a propagating electron resides in the surface, as opposed to the bulk.  The surface-bulk tunneling induces a pairing term $\tilde\Delta\tau_1$ in the surface where: 
\begin{equation} \tilde{\Delta} = (1-Z_\Gamma)\Delta_0 \end{equation} 
For strong surface-bulk coupling ($\gamma\gg\Delta_0$ or equivalently $Z_\Gamma\ll 1$) a sizeable fraction of the bulk pairing is induced on the surface. 

However, this is not the only effect of the interface.  From Eq. \ref{eq:SurfaceGrnsFn} we see that the surface-bulk coupling renormalizes the surface Hamiltonian, effectively rescaling the coefficients by a factor of $Z_\Gamma$:
\begin{equation} \mathcal{H}_{\text{TI/R}}\rightarrow \tilde{\mathcal{H}}_{\text{TI/R}} = Z_\Gamma\mathcal{H}_{\text{TI/R}}  \end{equation}
The effects of this renormalization are markedly different for topological insulators and Rashba 2DEG's, and we will consider each case in turn.

\subsubsection{Proximity Effect for TI Surface}
Renormalization of the topological insulator surface due to coupling to a bulk superconductor simply results in rescaling the Fermi velocity $v_F\rightarrow \tilde v_F=Z_\Gamma v_F$ and chemical potential $\mu\rightarrow\tilde{\mu}= Z_\Gamma\mu$. 
However, the nature of the induced pairing is independent of $v_F$ and $|\mu|$, rather, it depends only on the helical spin-winding of the surface Bloch-wavefunctions as one traverses a loop around the single Dirac cone in the surface Brillouin zone.
Since this helical winding is unchanged by the bulk-surface coupling, the induced pairing symmetry will be preserved regardless of $\gamma$.

Therefore, for TI materials there are no fundamental restrictions to pursuing arbitrarily strong coupling between the TI surface and the nearby bulk-superconductor, and consequently it is in principle possible to induce the full bulk-gap $\Delta_0$ on the TI surface.  This advantageous feature of TI surfaces was first pointed out in Ref. \onlinecite{TIInducedSC}.  In the subsequent section, we will see things are not so simple for the inducing superconductivity in a Rashba coupled 2DEG.

\subsubsection{Proximity Effect in a Rashba Coupled 2DEG}
Inducing superconductivity in Rashba coupled surface states is a more delicate matter than in TI surfaces.  Whereas the helical TI surface states are topologically guaranteed to convert induced s-wave pairing into effective $p+ip$ pairing, constructing effective $p+ip$ pairing in a Rashba coupled 2DEG requires a careful balance of spin-orbit coupling to create helical winding and magnetization $V_z$ to remove one of the helicities.  In particular, the Zeeman gap at $k=0$ is given by $|\tilde{V}_z-\tilde{\Delta}|$, and must not close.  Consequently, in order to engineer a topological superconductor, one must arrange for the surface magnetization to exceed the induced pairing amplitude: $\tilde V_Z>\tilde\Delta$.  Since $\tilde V_z = Z_\Gamma V_z$, and $\tilde\Delta = (1-Z_\Gamma)\Delta_0$, one immediately sees that a balance must be struck to ensure that the bulk--surface interface is sufficiently strong to induce pairing, but not so strong that it destroys the magnetization gap at $k=0$.

\begin{figure}[ttt]
\begin{center}
a)\includegraphics[width=1.5in]{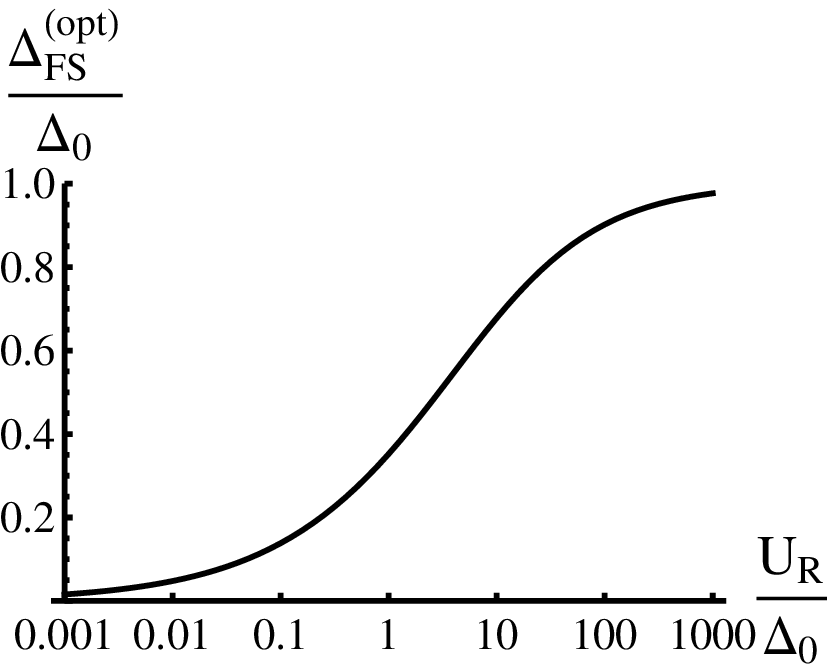}
b)\includegraphics[width=1.5in]{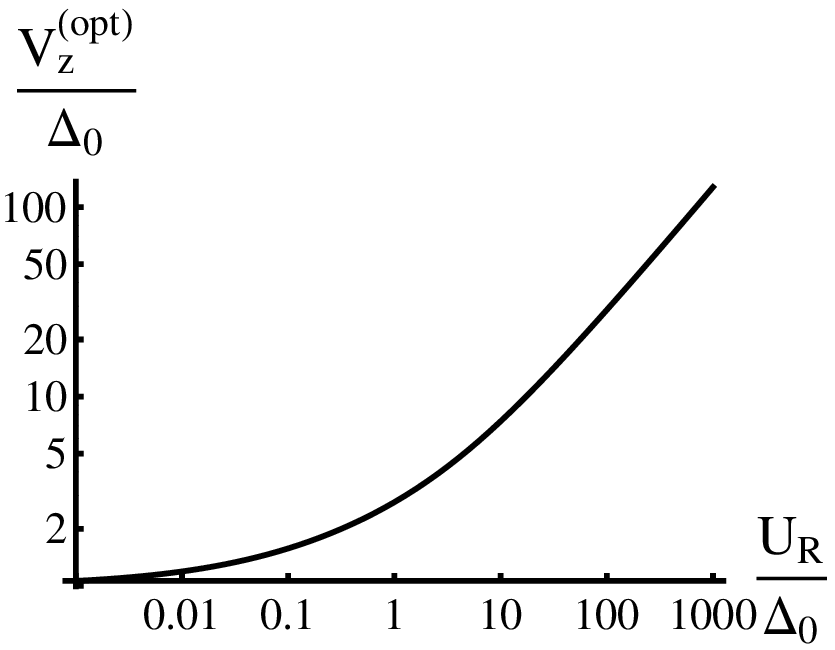}
c)\includegraphics[width=1.6in]{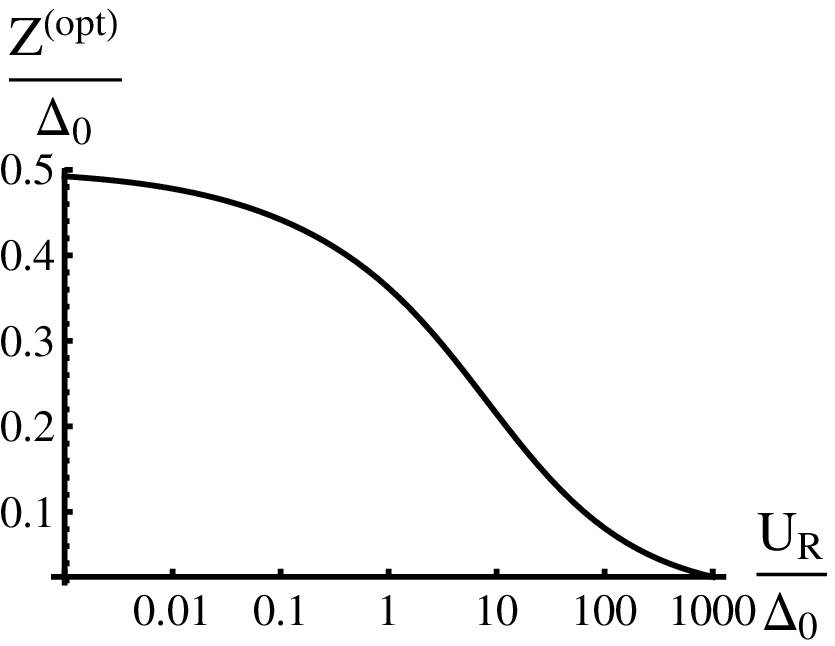}
d)\includegraphics[width=1.5in]{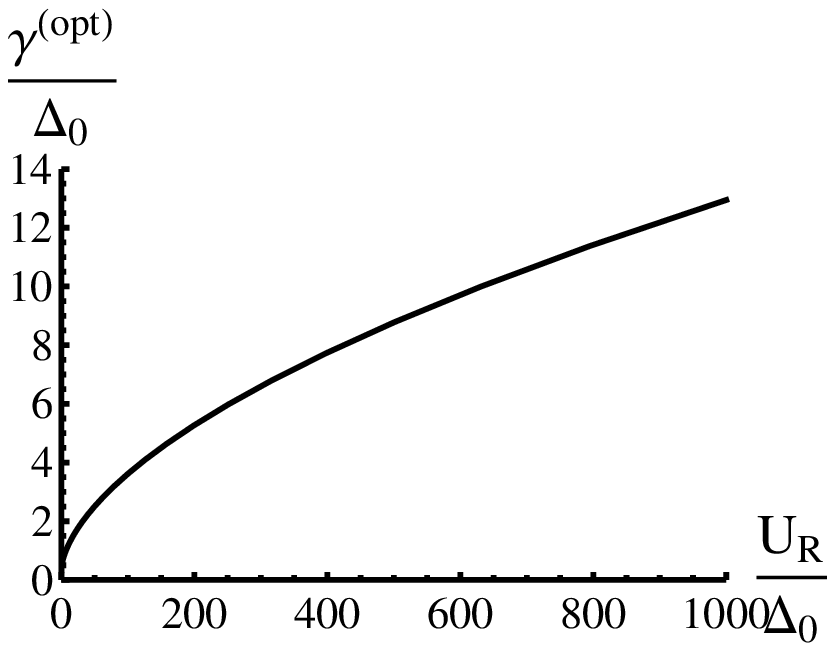}
\end{center}
\caption{Optimal parameters for producing a large pairing gap in a Rashba coupled surface via the proximity effect as a function of Rashba coupling strenght $U_R$.  Panels a.--d. respectively show the optimal excitation gap $\Delta_{\text{FS}}^{(\text{opt})}$, Zeeman splitting $V_z^{(\text{opt})}$, quasi-particle residue $Z^{(\text{opt})}$, and surface-superconductor coupling energy $\gamma^{(\text{opt})}$.  All energies are measured with respect to the pairing amplitude $\Delta_0$ of the bulk superconductor.}
\label{fig:OptimalInducedGap}
\end{figure}

We now turn to a quantitative analysis of the effect of proximity induced pairing in a Rashba 2DEG.  The goal will be to find the optimum set of parameters in order to achieve  a large p-wave superconducting gap on the Rashba coupled surface.  There are two excitation gaps, the magnetization gap, $|\tilde{V}_z-\tilde\Delta| = |Z_\Gamma V_z-(1-Z_\Gamma)\Delta_0|$, at $k=0$ and the p-wave pairing gap $\tilde\Delta_{FS} = (1-Z_\Gamma)\Delta_{FS}$ at the Fermi-surface.  The smaller of these two energy scales sets the minimum excitation energy gap.  The former decreases with $Z_\Gamma$, whereas the latter increases with $Z_\Gamma$, indicating that the optimal $Z_\Gamma$ is:
\begin{equation} Z^{(\text{opt})} = \frac{\Delta_{FS}+\Delta_0}{V_z+\Delta_{FS}+\Delta_0} \end{equation} 
which corresponds to an optimal excitation gap of:
\begin{equation} \Delta_{\text{FS}}^{(\text{opt})} = \frac{V_z}{V_z+\Delta_{FS}+\Delta_0}\Delta_{FS} \end{equation}
where $\Delta_{FS}$ is a function of $U_R$, $V_z$, and $\Delta_0$ given by Eq. \ref{eq:DeltaFS}. 

Figure \ref{fig:OptimalInducedGap}a. shows the optimum achievable value of $E_{\text{gap}}$ as a function of Rashba coupling strength $U_R$, and Figure \ref{fig:OptimalInducedGap}b--d. show the corresponding optimal values of $V_z$, $Z_\Gamma$, and  $\gamma$.  In practice, it will likely not be possible to fine-tune the interface transparency, $\gamma$, between the Rashba surface and the adjacent superconducting layer, or the induced magnetization $V_z$.  Rather, these parameters will be determined by the detailed structure of the surface-superconductor and surface-magnetic insulator interfaces.  Therefore, one should view Figure \ref{fig:OptimalInducedGap}a. as an upper bound on the practically achievable induced pairing gap.  Even so, the general trend is clear: \emph{for small Rashba splitting, $U_R$, only a small fraction of the bulk pairing gap $\Delta_0$ is induced on the surface, whereas for large $U_R$ a substantial fraction of $\Delta_0$ is achievable}.

This analysis highlights one of the potentially serious drawbacks of using materials with weak spin-orbit coupling.  In the limiting case of small $U_R$, it is advantageous to arrange $Z_\Gamma\simeq 1/2$ , large $\Delta_0\gg U_R$, and $V_z = \Delta_0$, in which case:
\begin{equation} \lim_{\displaystyle U_R\rightarrow 0}E_{\text{gap}} \lesssim\frac{\sqrt{U_R\Delta_0}}{2} \end{equation} 
In particular for semiconductor materials in which typical Rashba couplings are typically of the order of $0.2-0.8K$\cite{AliceaSingle}), even after carefully optimizing $V_z$, $\gamma$, and $\Delta_0$ only a p-wave pairing gap on the order of $0.1-0.4K$ is achievable.  Such small excitation gaps would require operating at temperatures much smaller than $0.1K$ in order to avoid thermal excitations, which could pose difficulties for experiments.  Furthermore, as will be shown in more detail below, small $U_R$ puts stringent restrictions on sample purity, as even small amounts of disorder will further suppress induced pairing.  In contrast, the situation is much more hopefull for materials with strong spin orbit couplings.  For strong Rashba couplings, one induce nearly the full superconducting gap, $\Delta_0$, given sufficiently transparent superconductor-surface interfaces.  

\subsection{Surface Resonances}
So far, we have been implicitly considering an artificial interface between a 2D material (either a TI surface or Rashba 2DEG) and a different superconducting material.  A potentially simpler alternative for realizing $p+ip$ superconductivity, is to use the naturally occuring interface between a bulk-superconductor and its surface.  This approach would eliminate the need to find compatible materials to engineer an appropriately transparent interface.

The formalism developed above applies equally well in this case.  Namely, if electronic states on the surface of a bulk metal occur at the same energy and momentum as bulk states, then the surface states decay into the bulk leaving behind broadened resonances.  If the bulk becomes a superconductor, the surface-bulk coupling induces superconductivity on the surface.  Denoting the width of the surface-resonance (in the absence of bulk-superconductivity) by $\gamma$, the induced superconductivity is again described by Eqs. \ref{eq:SurfaceGrnsFn} and \ref{eq:SurfaceZ}.  It is also possible that surface states coexist at the same energy as bulk bands, but reside in regions of the Brillouin zone for which there are no bulk-states.  In this case there is no direct tunneling from the surface into the bulk, and the surface state would remain sharp state rather than broadening into a resonance. Consequently to obtain superconductivity on the surface, one would need to rely on some scattering process (e.g. phonon, electron-electron, or disorder scattering) to transfer electrons between surface and bulk states.  

For natural superconducting metals with strong spin-orbit coupling (such as Pb), the electrostatic potential created by the material's surface interupts the bulk inversion symmetry, giving rise to a surface Rashba coupling.  If the surface Hamiltonian has appropriate combinations of Rashba spin-orbit coupling and magnetization (as described above), then the induced surface superconductivity will again have effective $p+ip$ pairing symmetry.  

A related approach is possible for topological insulator materials, where it has been demonstrated\cite{TISC} that doping can produce superconductivity with transition temperatures $T_C\sim 0.15-5.5K$.  Furthermore, it is common\cite{Bergman} that samples of materials such as $\text{Bi}_2\text{Se}_3$ that are expected to be bulk-insulators, are actually metallic.  In these ``topological metals", the topologically protected surface states that would appear for a bulk insulator appear instead as resonances\cite{Bergman}.  In fact a large amount of experimental effort is currently focused on finding materials with genuinely insulating bulks in order to investigate surface electron transport.  However, for the purpose of engineering a $p+ip$ superconductor, this surface-bulk coexistence is actually advantageous, and the combination of bulk superconductivity and surface-bulk coupling will result in an effective $p+ip$ superconductor at the surface of a superconducting topological metal.  

To examine whether $p+ip$ superconductors built from surface resonances also exhibit Majorana bound states, for example in vortex cores or at the ends of one-dimensional magnetic domains, one can write down the $T$--matrix for scattering from a vortex or domain wall and look for poles at zero-energy.  For a static vortex or domain wall configuration, the $T$--matrix at zero-energy is constructed from various products of surface Green's functions (see Eq.\ref{eq:SurfaceGrnsFn}) also at zero-energy.  Since $\Sigma_\Gamma(\omega=0) = Z(\omega=0)\Delta_0\tau_1$, the surface Green's function is identical to that of an ideal $p+ip$ superconductor with gap $\Delta = Z\Delta_0$. \emph{Therefore surface-resonance $p+ip$ superconductors will exhibit zero-energy Majorana bound-states under exactly the same conditions as the effective $p+ip$ superconductor discussed previously}.  These Majorana states are localized to the surface layer and are protected against decaying into bulk states because of the bulk superconducting gap.

\section{Disorder}
In this section we show that disorder effects the induced superconductivity very differently in the TI scheme as opposed to the Rashba scheme.  To model disorder, we consider a random on-site potential 
\begin{equation} H_{Dis}=\sum_{r,\sigma} V(r)c_{r,\sigma}^\dagger c_{r,\sigma}  \end{equation}
 that has only short-range correlations
\begin{equation} \overline{V(r)V(r')} = W^2\delta(r-r')  \end{equation}
where $W$ is the disorder strength and $\overline{\(\cdots\)}$ indicates an average over disorder configurations.  It is useful to parameterize disorder either by the scattering time $\tau \equiv 1/N(0)W^2$ or the mean free path $\ell = v_F\tau$ where $v_F$ is the Fermi velocity of the surface layer and $N(0)$ is the surface--density of states. Furthermore, we consider moderate disorder that is too weak to induce localization, specifically that $k_F\ell\gg 1$, but make no other assumptions on disorder strength.  Since non-planar disorder scattering diagrams are sub-leading in $(k_F\ell)^{-1}$, the regime $k_F\ell\gg 1$ allows for a controlled expansion for the disorder self-energy.  

The disorder averaged self-energy and Green's function are related by the following set of coupled equations:
\begin{eqnarray} \overline{\mathcal{G}}(i\omega,k)&=& \[\mathcal{G}_0(i\omega,k)^{-1}-\Sigma(i\omega)\]^{-1}  \nonumber \\  
\Sigma(i\omega) &=& W^2\tau_3\sum_k \overline{\mathcal{G}}(i\omega,k) \tau_3 \label{eq:SelfConsEqns}
 \end{eqnarray}
where the bare (non-disordered Green's function) $\mathcal{G}_0$ is given by Eq. \ref{eq:SurfaceGrnsFn}, and incorporates the proximity induced superconductivity.

\begin{figure}[ttb]
\begin{center}
\includegraphics[width=3.2in]{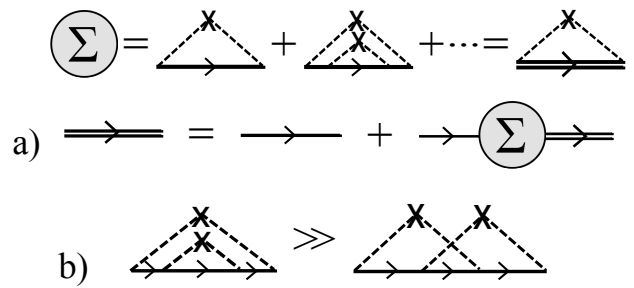}
\end{center}
\caption{Panel A shows a diagrammatic representation of Eq. \ref{eq:SelfConsEqns} for the disorder averaged Green's function and self-energy respectively.  Disorder scattering is represented by dashed line originating from an $\times$.  For delta-function-correlated impurities only multiple scatterings from the same impurity contribute.  Panel B shows an example of a crossed diagram (right) that is sub-leading in $(k_F\ell)^{-1}$ compared to the non-crossed diagram with the same number of disorder scatterings (left).}\label{fig:Diagrams}
\end{figure}

\subsection{Time-Reversal Symmetry}
In the subsequent discussion of disorder, the presence or absence of time-reversal (TR) symmetry plays a key role.  We will presently show that when (TR) symmetry is present, induced superconductivity is immune to the presence of disorder.  The proof of this principle is most conveniently conducted in the basis of time reversed pairs\cite{AndersonThm} (see Eq. \ref{eq:TRBasis}), in which the Hamiltonian for the disordered system with time-reversal symmetric s-wave pairing induced by proximity effect can be written as the following block-matrix:
\begin{equation} H = \begin{pmatrix} H_0+V&\Delta\mathbb{I} \\ \Delta\mathbb{I}&-(H_0+V) \end{pmatrix} \label{TRHBdG}\end{equation} 
Here $H_0$ is the Hamiltonian of the (clean) surface, $V$ is random on-site disorder, $\Delta$ is the induced pairing amplitude, $\mathbb{I}$ is the $N\times N$ identity matrix where $N$ is the number of degrees of freedom in the system.  Since $\Delta\mathbb{I}$ commutes with $H_0$ and $V$, the eigenvalue problem $\det\(H-\e\)=0$ can be simplified:
\begin{equation} 0=\det\(H-\e\) = \det\(\e^2\mathbb{I}-(H_0+V)^2-\Delta^2\)  \end{equation}
Denoting the eigenvalues of $H_0+V$ by $\{\tilde{\e}_n\}$, the eigenvalues of $H$ are $\pm \tilde{E}_n$ where:
\begin{equation} \tilde{E}_n = \sqrt{\tilde{\e}_n^2+\Delta^2}  \end{equation}
which is bounded below by $\Delta$, independent of the particular disorder configuration.  

These manipulations show that, so long as the 2D surface Hamiltonian is TR invariant, disorder cannot reduce the superconducting gap. As an aside, it is useful to note that the above considerations do not depend on $V$ being spin-independent so long as it preserves TRI.  In particular strong spin-orbit impurity scattering will also not reduce the superconducting gap.  


\subsection{Disordered Topological Insulators}
Since the TI Hamiltonian (Eq. \ref{eq:HTI}) is TR invariant, from the previous discussion we know that the pairing gap cannot be diminished by disorder.  As a simple demonstration of this general principle, one can explicitly calculate the disorder averaged Green's function and self-energy (see Eq. \ref{eq:SelfConsEqns}).  

In typical experimental situations, the TI surface states are intrinsically doped away from the surface Dirac point leaving an appreciable density of states at the Fermi surface\cite{Bergman}.  In this case, $\mu\gg\Delta$, and one finds:
\begin{equation} \Sigma(i\omega) = \cfrac{\tau^{-1}}{\sqrt{\Delta^2+\omega^2}}\(i\omega-\Delta\tau_1\)   \end{equation}
where $\tau^{-1} = \pi N(0)W^2$ is a measure of the disorder strength. Incorporating this disorder self-energy into the Green's function results in a disorder averaged Green's function of the same form as the bare Green's function $\mathcal{G}_0$, except with a reduced quasi-particle weight $Z$ due to disorder scattering:
\begin{eqnarray} \overline{\mathcal{G}}(i\omega) &=& \frac{Z}{i\omega-Z\mathcal{H}_{TI}-\Delta\tau_1}  \nonumber \\
Z(i\omega) &=&  \[1+\cfrac{\tau^{-1}}{\sqrt{\Delta^2+\omega^2}}\]^{-1} \hspace{.15in} (\mu\gg \Delta)  \label{eq:TIDisorderGrnFunction} \end{eqnarray}
For chemical potential tuned to the Dirac point, the results are similar except that the quasi-particle weight takes a different form: 
\begin{equation} Z(i\omega) = \[1+\cfrac{W^2}{2\pi v^2}\ln\(\cfrac{\Lambda^2}{\Delta^2+\omega^2}\)\]^{-1}  \hspace{.15in} (\mu=0)  \end{equation}
In either case, inspection of Eq. \ref{eq:TIDisorderGrnFunction} reveals that the minimum excitation gap, $\Delta$, is unchanged by disorder scattering, in agreement with the general principles outlined above for TR invariant systems.

\subsection{Disordered Rashba 2DEGs}

In contrast to the TI case, creating a topological superconductor from a Rashba 2DEG requires explicitly breaking TR symmetry by inducing surface magnetization $V_z$. Without TR symmetry, the general arguments outlined above do not apply, and the induced pairing is vulnerable to disorder scattering.  The analysis below demonstrates that the pair-breaking effects of disorder scattering are dramatically enhanced by the singular density of states at the superconducting gap edge, and furthermore that these effects are especially pronounced in systems with weak Rashba coupling.  Since the pairing on the surface is induced by the bulk, it never vanishes.  However, we shall see that pairing can be greatly reduced by even a small amount of disorder unless $U_R\gg V_z$.  In fact, when $U_R\ll V_z$, the suppression due to disorder is more severe than for conventional superconductors with magnetic impurities for which the superconducting gap typically closes for $\xi_0/\ell \sim 1$.  We will see, for $U_R\ll V_z$, that disorder strongly suppresses the induced superconductivity even for very weak disorder.


For superconductivity induced by proximity effect, the surface state wave-functions are localized to the surface, but extend into the superconductor with characteristic lengthscale $\xi_L$.  Therefore electrons residing in the Rashba material are scattered not only by impurities in the Rashba material and interface roughness, but also by impurities in the superconductor.  For example, even if one starts with a pristine semiconductor structure (such as a self-assembled nanowire), if superconductivity is induced by proximity to superconductor with some impurities then the mean-free path will be set by the superconductor rather than the semiconductor.  

To better understand the distinction between impurities residing in the Rashba coupled surface and those in the superconductor, we consider each separately.  To treat either case we find that it is sufficient to replace the bare disorder scattering time $\tau^{-1} = \pi N(0)W^2$ (where $N(0)$ is the surface density of states) by an effective disorder scattering time 
\begin{equation} \(\tau^{-1}\)_{\text{eff}} = \left\{\begin{array}{ll} Z_\Gamma^2\tau^{-1} &; \text{  Surface Disorder} \\ (1-Z_\Gamma)^2\tau^{-1} &; \text{  Bulk Disorder}  \end{array} \right. \label{eq:EffectiveScatteringTime}\end{equation}
An explicit derivation of these expressions is given in Appendix A, but the effective scattering time can be understood more simply as follows.  The fraction of the surface--resonance wave--function which lies on the surface is $Z_\Gamma$ whereas the fraction residing in the bulk superconductor is $(1-Z_\Gamma)$.  Therefore, the disorder scattering matrix elements should be weighted by either $Z_\Gamma$ or $(1-Z_\Gamma)$ for surface and bulk disorder respectively.  Eq. \ref{eq:EffectiveScatteringTime} can then be simply understood by noting that the scattering time is proportional to the square of the disorder matrix element. By inspection, we see that effects of surface disorder are suppressed in the limit of strong surface--bulk tunneling, whereas the effects of bulk disorder are suppressed in the limit of weak tunneling. Using this effective scattering time, we now turn to the problem of solving the self-consistency relations given in Eq. \ref{eq:SelfConsEqns} using the surface Green's function  in Eq. \ref{eq:SurfaceGrnsFn} which includes the effects of proximity to the bulk superconductor. 

For ordinary disordered superconductors, the strength of disorder is conveniently parametrized by the ratio of the coherence length $\xi_0 = \pi v_F/\Delta$ to the mean free path $\ell=v_F\tau$.  For the proximity induced superconductivity these parameters are renormalized by surface--bulk coupling and also depend on the type of disorder (surface or bulk).  We will see that the effects of disorder depends on disorder strength only through the ratio $\tilde{\xi_0}/\ell_{\text{eff}}$, where $\tilde{\xi_0} = \pi \tilde{v}_F/\tilde{\Delta}$ is the surface coherence length and $\ell_{\text{eff}}=\tilde{v}_F \tau_\text{eff}$ is the effective mean-free path for disorder electrons.  This effective ratio can be written in terms of the intrinsic ratio of the intrinsic mean-free path, $\ell$ (un-renormalized by proximity induced superconductivity) and coherence length of the bulk superconductor, $\xi_0$, as follows:
\begin{equation} \cfrac{\tilde{\xi_0}}{\ell_{\text{eff}}} = \left\{\begin{array}{ll} \displaystyle \frac{Z_\Gamma^2}{1-Z_\Gamma}\frac{\xi_0}{\ell} &; \text{  Surface Disorder} \\ \displaystyle (1-Z_\Gamma)\frac{\xi_0}{\ell} &; \text{  Bulk Disorder}  \end{array} \right. \label{eq:EffectiveDisorderRatio}\end{equation}
By working in terms of this effective ratio, it is possible to treat both the cases of surface--impurities and bulk--impurties on equal footing.

\subsubsection{Analytic Expressions for Weak Disorder}
For weak disorder ($\tilde{\xi_0}/\ell_{\text{eff}}\ll 1$), it is sufficient to evaluate the self-energy to lowest order in disorder scattering strength, corresponding to the first diagram for the self-energy shown in the top line of Fig. \ref{fig:Diagrams}:
\begin{equation} \Sigma^{(1)}(i\omega) = W^2\tau_3\sum_k \tilde{\mathcal{G}}_0(i\omega,k) \tau_3  \end{equation}
Here we emphasize that the value of $W^2$ should be appropriately renormalized according to Eq. \ref{eq:EffectiveScatteringTime} depending on whether the scattering considered occurs in the surface or in the bulk superconductor.  For $\tilde \Delta_{FS}\ll \tilde V_z$, the dominant contributions to the $k$-integral come from near the Fermi-surface.  Linearizing the Bugoliuobov dispersion about the Fermi-surface, and performing the integration yields:
\begin{equation} \Im m\Sigma^{(1)}(i\omega) \simeq -x\omega\[\frac{1}{2}+2\cfrac{\tilde V_z}{\tilde \Delta_{BG}^2}\sigma_z\(\tilde \Delta\tau_1-\tilde U_R\tau_3\)\]   \end{equation} 
\begin{eqnarray} \Re e\Sigma^{(1)}(i\omega) &\simeq& \frac{x}{\tilde \Delta_{BG}^2}\big{[}\(\omega^2-\tilde \Delta^2\)\tilde U_R\tau_3
 \\
&+&\tilde V_z\(\tilde \Delta^2+\omega^2\)\sigma_z+2\tilde U_R^2\tilde \Delta\tau_1\big{]}  \end{eqnarray}
\begin{equation} x = \cfrac{\pi N(0)W^2}{\sqrt{\tilde \Delta_{FS}^2+\omega^2}}\equiv \cfrac{\tau_\text{eff}^{-1}}{\sqrt{\tilde \Delta_{FS}^2+\omega^2}}  \end{equation}
where $\tau_\text{eff}^{-1}$ is a measure of the disorder strength, given by Eq. \ref{eq:EffectiveScatteringTime}.

This self-energy alters the spectrum of the disorder averaged BdG Hamiltonian.  For weak disorder, we expect the gap at the Fermi surface to change only slightly.  To find the correction to $\Delta_{FS}$ due to disorder, one needs to analytically continue the self-energy to real frequency, and then look for a pole in the disorder averaged Green's function at $\omega = \Delta_{FS}-\delta\omega$, i.e. to solve \begin{equation} 0 = \det\[\tilde \Delta_{FS}-\tilde \delta\omega - \mathcal{H}(k_F)-\Sigma^{(1)}\(\omega=\tilde \Delta_{FS}-\delta\omega\)\] \end{equation}  to leading order in $\delta\omega$ one finds: 
\begin{equation} \delta\omega = \Psi_0^\dagger \Sigma^{(1)}\(\omega=\tilde \Delta_{FS}-\delta\omega\)\Psi_0 \end{equation}
where $\Psi_0 = \begin{pmatrix}u_\up&u_\down&v_\down&-v_\up\end{pmatrix}^T$ is the eigenvector of $\mathcal{H}(k_F)$ with energy $\tilde \Delta_{FS}$. 

In the limiting case where $\tilde V_z\gg \tilde U_R$, $\Psi_0\simeq \frac{1}{\sqrt{2}}\begin{pmatrix}0&1&0&1\end{pmatrix}^T$, and one finds 
\begin{equation} \delta\omega \simeq \frac{\tilde \Delta^2}{4V_z}x \simeq \cfrac{\tilde \Delta^2 \tau_\text{eff}^{-1}}{4\tilde V_z\sqrt{2\tilde \Delta_{FS}\delta\omega}}  \end{equation}
Using $\Delta_{FS}\simeq \Delta\sqrt{\cfrac{U_R}{V_z}}$, and solving for $\delta\omega$ gives the following expression for the disorder renormalized pairing gap at the Fermi-surface:
\begin{equation} \tilde{\Delta}_{FS}(\tau_\text{eff}^{-1}) \simeq \tilde \Delta\sqrt{\cfrac{\tilde U_R}{\tilde V_z}}\[1-\(\cfrac{\tilde V_z}{4\sqrt{2}\tilde U_R }\cfrac{\tilde{\xi_0}}{\ell_{\text{eff}}}\)^{2/3}\] \hspace{.1in}(\tilde V_z\gg \tilde U_R) \label{eq:RashbaDisVz} \end{equation}  
The unusual non-analytic dependence on disorder strength stems from the singular behavior of $x$ as $\omega\rightarrow \tilde \Delta_{FS}$, which in turn reflects the Van-Hove singularity in the superconducting density of states at the gap edge.  This Van-Hove singularity enhances the effective disorder strength $x$, and in particular leads to an infinite slope of $\tilde{\Delta}_{FS}(\tau_\text{eff}^{-1})$ as $\tau_\text{eff}^{-1}\rightarrow 0$.  

In the opposite limit, where $\tilde U_R\gg \tilde V_z$, $\Psi_0\simeq \frac{1}{2}\begin{pmatrix}1&-1&1&-1\end{pmatrix}^T$ and consequently the weak disorder correction to the gap energy vanishes to leading order.  Including sub-leading contributions in $\tilde U_R/\tilde V_z$ results in:
\begin{equation} \tilde \Delta_{FS}(\tau_\text{eff}^{-1})\simeq\tilde \Delta\[1-\(\cfrac{18}{\sqrt{2}}\cfrac{\tilde V_z^2}{\tilde U_R^2}\cfrac{\tilde{\xi_0}}{\ell_{\text{eff}}}\)^{2/3}\] \hspace{.15in}(\tilde U_R\gg \tilde Vz)  \label{eq:RashbaDisUr}\end{equation}

\begin{figure}ttt]
\begin{center}
\includegraphics[width=3.2in]{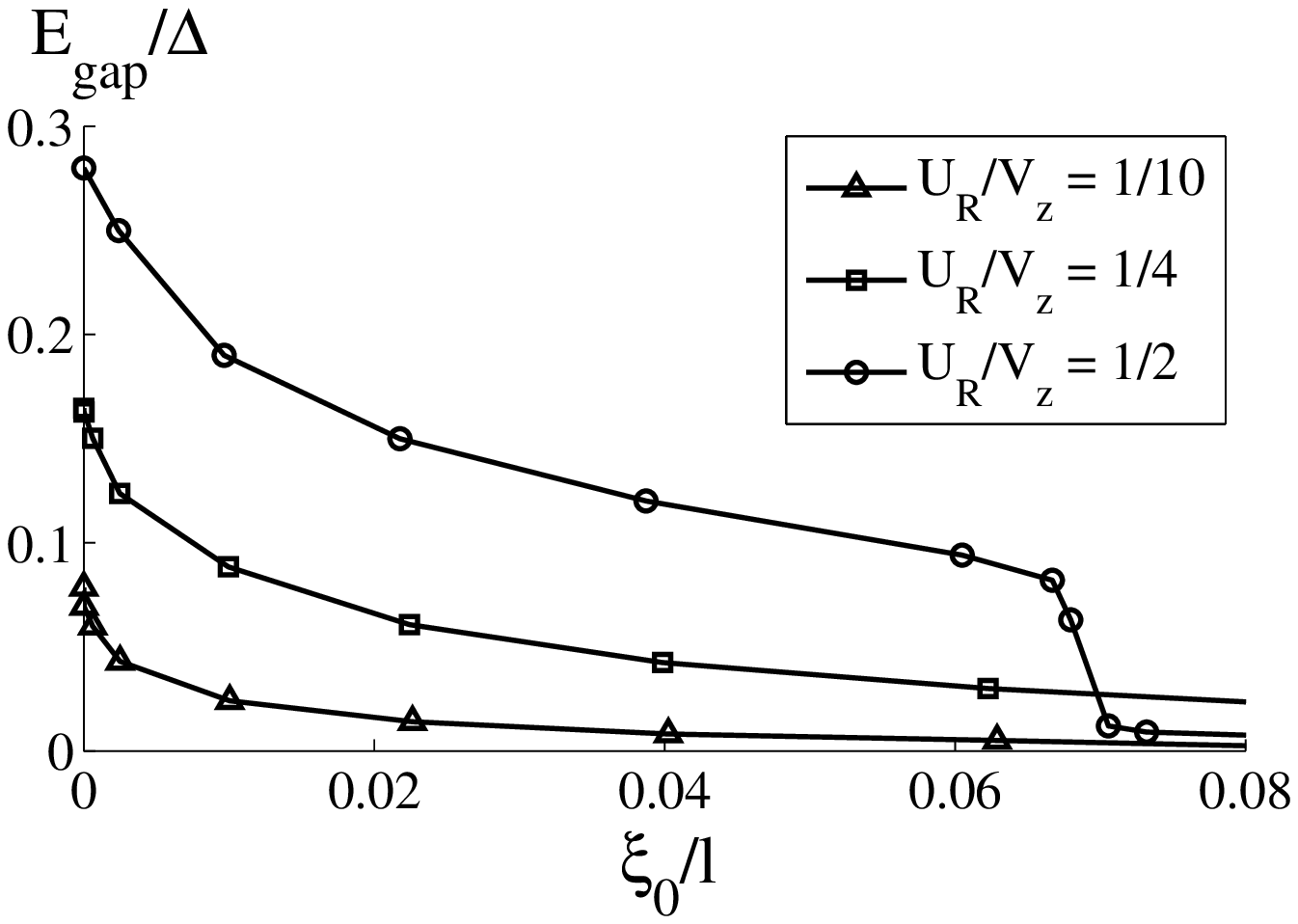}
\end{center}
\vspace{-.2in}
\begin{center}
\includegraphics[width=3.2in]{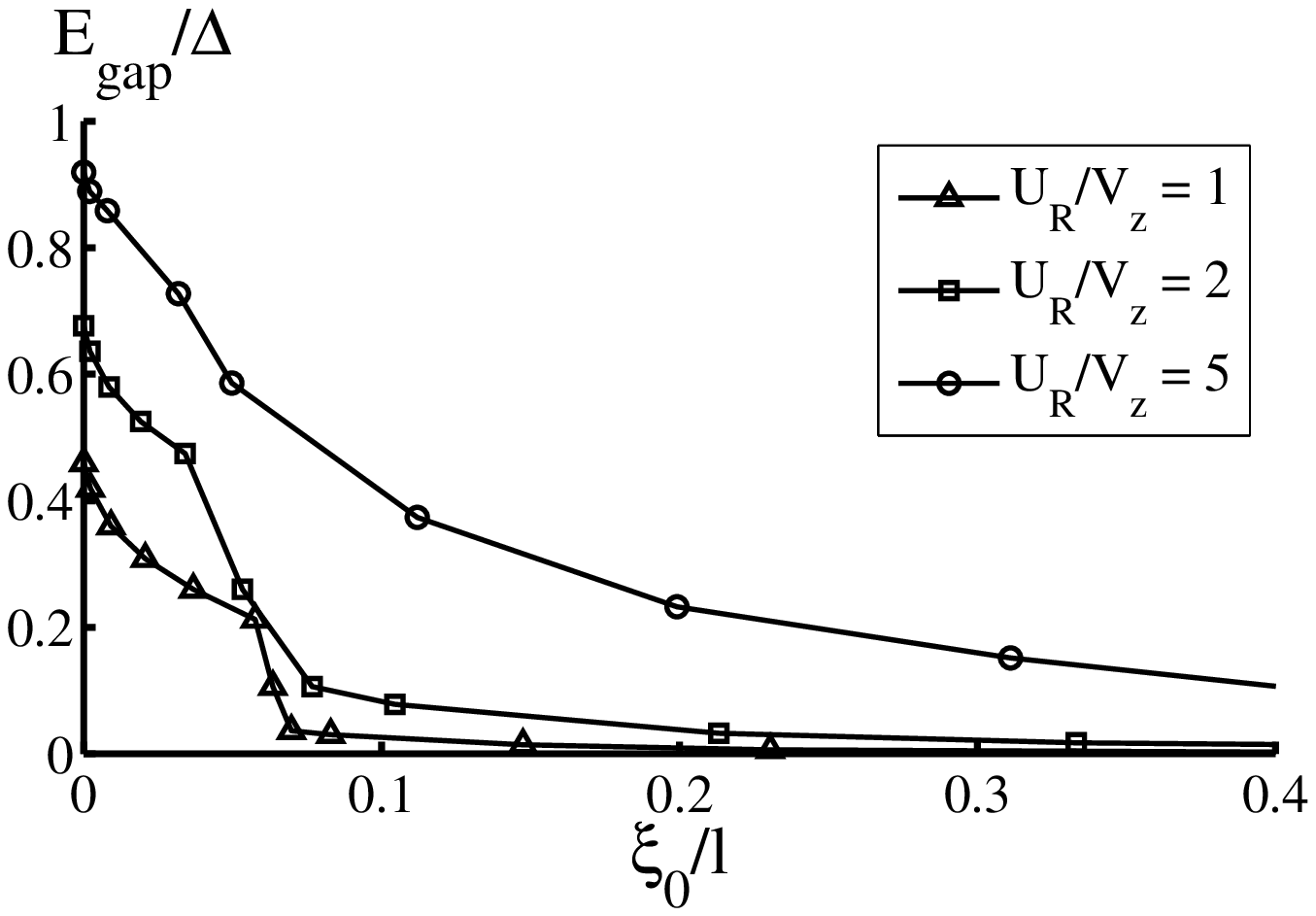}
\end{center}
\vspace{-.2in}
\caption{The excitation gap $E_{\text{gap}}$ as a function of coherence length $\tilde{\xi}_0=\pi v_F/\tilde \Delta$ to the effective mean free path $\ell_{\text{eff}} = \tilde{v}_F\tau_{\text{eff}}^{-1}$. $E_{\text{gap}}$  is obtained from numerically solving Eq. \ref{eq:SelfConsEqns} for a Rashba 2DEG with induced magnetization $\tilde V_z$ and superconductivity $\tilde \Delta$.  Here $\widetilde{\(\cdots\)}$ indicates renormalization due to the proximity effect. The effective disorder strength has a different form depending on whether disorder scattering occurs predominantly in the surface-layer or in the bulk superconductor.  Both cases can be treated by choosing the appropriate expression for $\tilde{\xi}_0/\ell_{\text{eff}}$ from Eq. \ref{eq:EffectiveDisorderRatio}. The parameters used in this simulation were $t=1$, $\tilde V_z = 0.1$, $\tilde \Delta = 0.01$, and various values of $\tilde U_R$. The top panel shows curves for $\tilde V_z\gg \tilde U_R$, the regime appropriate for semiconductor materials, whereas the bottom panel shows curves in the $\tilde U_R\gtrsim \tilde V_z$ regime which could be achieved by using metallic thin films with stronger spin-orbit coupling. The magnetization $\tilde V_z$ breaks time reversal symmetry rendering the induced pairing susceptible to disorder.  For $\tilde V_z\gg \tilde U_R$ the gap is already strongly supressed when $\tilde{\xi}_0$ is only a few percent of $\ell_{\text{eff}}$.\label{fig:NumericalDisorder}}
\end{figure}

\subsubsection{Numerical Solution for Moderate Disorder}  
For stronger disorder, Eq. \ref{eq:SelfConsEqns} must be solved self-consistently, which can be done numerically.  In order to regulate the numerical integrals in the UV we replace the continuum dispersion with a periodic one of the form $\xi_k = -2t\cos(k)$ which naturally introduces a finite band-width.  The top and bottom panels of Figure \ref{fig:NumericalDisorder} show the dependence of the induced superconducting gap on disorder strength for $\tilde U_R\gtrsim \tilde V_z$ and $\tilde V_z\gg \tilde U_R$ respectively.

For very weak disorder, $\tilde{\xi}_0\ll \ell_{\text{eff}}$, the excitation gap exhibits non-analytic infinite initial slope predicted by Equations \ref{eq:RashbaDisVz} and \ref{eq:RashbaDisUr}.  Stronger disorder never fully closes the superconducting gap, however for $\tilde V_z\gg \tilde U_R$, the gap is largely suppressed even when $\tilde{\xi}_0$ is only a few percent of $\ell_{\text{eff}}$.  In most cases, $E_{\text{gap}}$ is suppressed smoothly with increasing disorder strength, however, the $E_{\text{gap}}/\tilde \Delta$ curves for $\tilde U_R\simeq \tilde V_z$ have a knee-shaped kink at $\tilde{\xi}_0/\ell_{\text{eff}}\simeq 0.07$ after which $E_{\text{gap}}$ drops abruptly.  This knee occurs when disorder reduces of the magnetization gap at $k=0$ below the pairing gap $\tilde \Delta_{FS}$ at the Fermi-surface. 

These results can be readily applied both to the case where disorder scattering is due to the adjacent superconductor, and when disorder scattering occurs in the surface material, by choosing the form for $\tilde{\xi}_0/\ell_{\text{eff}}$ from Eq. \ref{eq:EffectiveDisorderRatio}.  It is possible to reduce the sensitivity to bulk--disorder by reducing the surface--bulk tunneling strength, $\Gamma$.  However, reducing $\Gamma$ will also cause a smaller proximity induced pairing gap $\tilde{\Delta}$.  Similarly, it is possible to reduce the sensitivity to surface disorder by increasing the surface--bulk tunneling rate, though, doing so will be detrimental if the surface layer is cleaner than the bulk superconductor.

Here we see a second drawback of using materials with low Rashba coupling: in addition to limiting the size of the induced pairing gap in the absence of disorder, small Rashba coupling renders the topological superconductor susceptible even to small amounts of disorder ($\tilde{\xi}_0/\ell_{\text{eff}} \ll 1$).  While bulk semiconductors are typically cleaner than metallic thin films, their extreme sensitivity to disorder will likely be problematic.  In particular, great care would need to be taken to limit interfacial roughness between the semiconductor and adjacent bulk superconductor and magnetic insulating film.

Before concluding, we remark on two possible extensions of this analysis.  Firstly, the effects of disorder were treated for fully two-dimensional structures, whereas Majorana fermions emerge in one-dimensional (or quasi-one-dimensional) geometries.  The effects of disorder in quasi-one-dimensional Rashba coupled structures were analyzed numerically in Ref. \onlinecite{Potter1,Potter2}, and give similar results to those given above for two-dimensions. Finally, while this analysis has been carried out for the case of Rashba-type spin-orbit coupling, we expect similar results for systems in which both Rashba and Dresselhaus-type spin-orbit couplings are present.  The relevant factor in either case is the presence of magnetization $V_z$ which breaks time-reversal symmetry and renders the induced superconductivity susceptible to disorder regardless of spin-orbit type.

\section{Discussion and Conclusion}
In conclusion, we have compared the prospects for constructing an effective $p+ip$ superconductor from TI and Rashba 2DEG based materials.  We have focused on technical limitations to inducing superconductivity in these materials, and examined the effects of disorder on the induced superconductivity.  In both regards, the TI materials offer natural advantages.  In particular, the effective $p+ip$ nature of induced superconductivity in a TI surface is guaranteed by the intrinsic spin-helicity of the bare TI surface states.  As a consequence, there are no fundamental limitations for inducing superconductivity by the proximity effect.  Furthermore, since time-reversal symmetry remains intact in the TI surface, the induced superconductivity is guaranteed, based on general principles, to be immune to disorder.

While TI surface states offer certain advantages, TI materials are relatively new and  many materials challenges remain. It may therefore be desireable to construct a $p+ip$ superconductor from more conventional materials with strong Rashba spin-orbit coupling.  Here one needs to induce superconductivity by proximity to a conventional superconductor, and to induce magnetization for example by proximity to a ferromagnetic insulator.  In this case one must strike a comparatively delicate balance of spin-orbit coupling and induced magnetization to ensure that the resulting superconductor is effectively $p+ip$\cite{AliceaSingle}.  

In this regard, materials with small Rashba coupling strengths face serious difficulties: 1) the size of the induced superconducting gap is limited by the size of the Rashba coupling and 2) for weak Rashba coupling the induced superconductivity is quite fragile and is strongly suppressed even by small amounts of disorder.  Our analysis indicates that an alternative class of materials with stronger spin-orbit coupling should be sought.  For example, metallic thin films with heavy atomic elements can give orders of magnitude larger spin-orbit couplings than the semiconductor materials that have so far dominated the theoretical discussion.  

Constructing a $p+ip$ superconductor from a Rashba 2DEG in this way requires engineering a complicated set of material interfaces between the Rashba 2DEG, superconductor, ferromagnetic insulator, and gate electrodes.  However, a fortuitous choice of material could obviate the need for the superconducting and ferromagnetic interfaces.  For example, one could try to find bulk metals with strong bulk spin-orbit coupling that naturally superconduct and possess surface resonances.  In such a scenario, the coupling between surface-and bulk will then automatically induce superconductivity on the surface, eliminating the need to build an artificial interface for this purpose.  A further simplification is possible the metallic surface which has the appropriate symmetry such that both Rashba and Dresselhaus type spin-orbit coupling is present\cite{AliceaSingle}.  In this case, one could induce magnetization by applying an external field rather than by depositing a ferromagnetic insulator\cite{AliceaSingle}.  If a naturally superconducting material can be found with the appropriate spin-orbit coupled surface resonances, this approach might offer the simplest route to an artificial $p+ip$ superconductor and Majorana fermions.  

\textit{Acknowledgements - } We thank T. Oguchi for helpful discussion and for sharing data from first-principle calculations of the Au(110) surface band-structure. This work was supported by DOE Grant No. DE--FG02--03ER46076 (PAL) and NSF IGERT Grant No. DGE-0801525 (ACP).  Shortly after this work was submitted, another paper appeared that also analyzes the effects of disorder on superconductivity induced in one-dimensional Rashba coupled nanowires\cite{OtherDisorderPaper}.

\appendix
\section{Surface vs. Bulk Disorder}
In the main text, we argued on conceptual grounds, that the above analysis for disorder is easily modified to separately treat the two distinct cases where disorder scattering occurs predominantly in the surface layer or the adjacent bulk superconductor, by replacing the bare disorder scattering time by an effective scattering time weighted by $Z_\Gamma^2$ or $(1-Z_\Gamma)^2$ respectively.  Here we consider each case separately and provide an explicit demonstration of this claim.

\subsection{Surface Disorder}
Consider first the case of a pristine superconductor so that impurity scattering occurs only in the surface layer.  In this case the disorder scattering matrix elements are proportional to the fraction of the electron wave-functions that resides on the surface.  More precisely, we can calculate self-energy from disorder scattering in Eq.  \ref{eq:SelfConsEqns} using the surface Green's function in Eq. \ref{eq:SurfaceGrnsFn}.  Equivalently, in Fig. \ref{fig:Diagrams}, one should take single solid lines to be the surface Green's function in Eq. \ref{eq:SurfaceGrnsFn}. By inspection, we see that the effective disorder matrix elements are reduced by a factor of $Z_\Gamma<1$.  Therefore, the disorder self-energy for surface-disorder is suppressed by a factor of $Z_\Gamma^2$.

The analysis in the main text can be easily modified to treat the case where disorder occurs predominantly in the surface layer by replacing $\tau_1$ for the surface (without surface--bulk tunneling) by the effective ratio:
\begin{equation} \tau_{\text{eff}}^{-1} \equiv Z_\Gamma^2\tau^{-1}\end{equation} 
and keeping the same ratio $U_R/V_z$ (since this ratio is unaffected by the surface--bulk coupling).

\subsection{Bulk Disorder}
When the dominant source of disorder is in the adjacent bulk--superconductor, the surface electrons must first tunnel into the bulk in order to scatter from the disorder potential.  This leads to an renormalized effective disorder strength which is different than the bulk value.  To demonstrate this, we focus on the limit where surface--bulk tunneling is much stronger than bulk disorder ($\gamma\ll W$), in which case there is typically no more than one disorder scattering event per surface--bulk tunneling event.  

\begin{figure}[hhh]
\begin{center}
\includegraphics[width=1.5in]{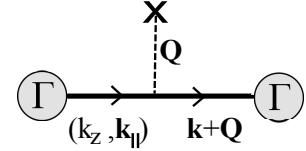}
\end{center}
\caption{Effective disorder vertex for surface--electrons in the case where disorder scattering occurs predominantly in the adjacent bulk--superconductor.  The solid lines are bulk Green's functions, circled $\Gamma$s indicate surface--bulk tunneling events, and the dashed line ending in $\times$ indicates disorder scattering that transfers momentum $\mathbf{Q}$}
\label{fig:DressedDisorderVertex}
\end{figure}

In this limit, it suffices to compute Eq. \ref{eq:SelfConsEqns} using the surface-Green's function in Eq. \ref{eq:SurfaceGrnsFn}, with an  effective disorder vertex is given by the diagram shown in Fig. \ref{fig:DressedDisorderVertex}.  In this figure the circled $\Gamma$ indicates surface--bulk tunneling and the dashed line indicates scattering from disorder. Written in terms of the bulk Green's functions the effective disorder vertex for surface states is:
\begin{eqnarray} &&\tilde{V}\(i\omega,\mathbf{k}_\parallel,\mathbf{Q}\)\tau_3 \nonumber \\ && = V|\Gamma|^2\sum_{ k_z} \cfrac{1}{i\omega-\xi_{(k_z,\mathbf{k}_\parallel)}\tau_3-\Delta\tau_1}\tau_3\cfrac{1}{i\omega-\xi_{\mathbf{k}+\mathbf{Q}}\tau_3-\Delta\tau_1}\nonumber\\ \end{eqnarray}
We are interested primarily in low-frequency behavior for which the external momenta $\{\mathbf{k}_\parallel, \mathbf{k}_\parallel + \mathbf{Q}_\parallel\}$ lie near the surface Fermi-level.  For a given $\mathbf{Q}_\parallel$ connecting two points on the surface Fermi-surface, there are two values of $Q_z$ for which $\mathbf{k}$ and $\mathbf{k}+\mathbf{Q}$ also lie on the surface.  Since the dominant contributions are for momenta lying near the Fermi-surface, we linearize the bulk-dispersion in the z-direction, and denote the tunneling density of states by $N_B(0)\equiv \(\cfrac{\partial \e_B(\mathbf{k})}{\partial{k_z}}\)^{-1}$.  With these approximations, the effective disorder potential for surface electrons is independent of momentum transfer $\mathbf{Q}_\parallel$:
\begin{equation} \tilde{V}(i\omega)\simeq \cfrac{\pi N_B(0)|\Gamma|^2}{\sqrt{\Delta^2+\omega^2}} \label{eq:EffectiveDisorderVertex}\end{equation}

The disorder self-energy given by the diagram in Fig. \ref{fig:Diagrams}a. can then be computed using the surface Green's function from Eq.\ref{eq:SelfConsEqns} and the effective disorder strength in Eq. \ref{eq:EffectiveDisorderVertex}.  When this self-energy is incorporated into the surface Green's function, it comes with a factor of $\(Z_\Gamma\)^2$.  Since we are interested in the behavior for $\omega\ll \Delta$, we may neglect the frequency dependence of various quantities, and we find that the effective disorder scattering time for surface-electrons is weighted (compared to the bulk quantity) by a factor:
\begin{eqnarray} \(\tau^{-1}\)_{\text{eff}} &=& \(\cfrac{\pi N_B(0)|\Gamma|^2}{\Delta+\pi N_B(0)|\Gamma|^2}\)^2 \(\tau^{-1}\)_{\text{bulk}} \nonumber \\
&=&(1-Z_\Gamma)^2\(\tau^{-1}\)_{\text{bulk}} \end{eqnarray}
For a transparent surface--bulk interface ($\pi N_B(0)|\Gamma|^2\gg \Delta$), this factor approaches unity, justifying the claim made in the main text.

\end{document}